\documentclass[onecollarge]{svjour3}       
       
\smartqed  
\usepackage{psfrag,graphicx,epsfig,color}
\usepackage{graphicx}
\usepackage{color}
\usepackage{epsfig}
\usepackage{subfigure}
\usepackage[dvipsnames]{xcolor}

\usepackage{epstopdf}

\usepackage{tikz}
\usetikzlibrary{arrows,intersections}
\usepackage{pgfplots}
\usepackage[all]{xy}

\begin{document}

\title{document}
\title{Flooding dynamics of diffusive dispersion in a random potential}

\author{Michael Wilkinson, Marc Pradas and Gerhard Kling}

\institute{Michael Wilkinson \at
             Chan Zuckerberg Biohub,\\ 
             499 Illinois Street,\\
             San Francisco,\\ 
             CA93107, USA\\
             \email{michael.wilkinson@czbiohub.org}      \\
             \\
             Marc Pradas \at 
             School of Mathematics and Statistics,\\
             The Open University,\\ 
             Walton Hall,\\ 
             Milton Keynes, MK7 6AA,\\
             England \\
             \email{marc.pradas@open.ac.uk}      \\
             \\
             Gerhard Kling \at
             University of Aberdeen,\\ 
             Business School, King's College,\\ 
             Aberdeen AB24 3FX,\\
             Scotland\\
             \email{gerhard.kling@abdn.ac.uk}}

\date{}

\maketitle

\begin{abstract}

We discuss the combined effects of overdamped motion in a quenched random potential 
and diffusion, in one dimension, in the limit where the diffusion coefficient is small. 
Our analysis considers the statistics of the mean 
first-passage time $T(x)$ to reach position $x$, arising from different realisations 
of the random potential: specifically, we contrast the median $\bar T(x)$, which is 
an informative description of the typical course of the dispersion, with the expectation value 
$\langle T(x)\rangle$, which is dominated by rare events where there is an exceptionally high
barrier to diffusion. We show that at relatively short times the median $\bar T(x)$ is explained 
by a \lq flooding' model, where $T(x)$ is predominantly determined by the highest barriers which 
is encountered before reaching position $x$. These highest barriers are quantified using methods of 
extreme value statistics.


\end{abstract}

\section{Introduction}
\label{sec: 1}

There are many situations where particles move under the combined influence of 
thermal diffusion and a static (or quenched) random potential \cite{Havlin2002}. The particles 
might be electrons, holes or excitons diffusing in a disordered metallic or 
semiconductor sample \cite{Cha+95}, or molecules diffusing in a complex environment such as the cytoplasm 
of a eukaryotic cell \cite{Saxton2007}. The state of knowledge of this problem 
is surprisingly under-developed, and in this work we present new results on 
the simplest version of this problem, in one dimension, where the equation of motion is
\begin{equation}
\label{eq: 1.1}
\dot x=-\frac{{\rm d}V}{{\rm d}x}
+\sqrt{2D}\,\eta (t)
\ .
\end{equation}
Here $V(x)$ is a random potential, $D$ is the diffusion coefficient, and $\eta(t)$ 
is a white noise signal with statistics defined by 
\begin{equation}
\label{eq: 1.2}
\langle{\eta(t)}\rangle=0
\ ,\ \ \ 
\langle\eta(t)\eta(t')\rangle=\delta (t-t')
\end{equation}
($\langle \cdot \rangle$ denotes expectation value throughout).
We assume that $V(x)$ is a smooth random function, defined by its statistical properties, which are 
stationary in $x$, and independent of the temporal white noise $\eta(t)$. 
The one and two-point statistics of this potential are 
\begin{equation}
\label{eq: 1.3}
\langle V(x)\rangle=0
\ ,\ \ \ 
\langle V(x)V(x')\rangle=C(x-x')
\ .
\end{equation}
The correlation function $C(\Delta x)$ is assumed to decay rapidly as $|\Delta x|\to \infty$. 
We also assume that the tails of the distribution of $V$ are 
characterised by a large-deviation \lq rate' (or \lq entropy') function $J(V)$, so that when $|V|$ is large,
the probability density function of $V$ is approximated by 
\begin{equation}
\label{eq: 1.4}
P_V\sim \exp[-J(V)]
\end{equation}
where throughout we shall use $P_X$ to denote the probability density function (PDF)
of a random variable $X$. If $P_V$ is a Gaussian distribution, then the 
entropy function is quadratic, $J(V)\sim V^2/2C(0)$.

It has been proposed that the behaviour of this system is diffusive, with 
an effective diffusion coefficient which vanishes very rapidly as $D\to 0$: 
Zwanzig  \cite{Zwanzig1988} gave an elegant argument which implies that,
when $V(x)$ has a Gaussian distribution, the effective diffusion coefficient 
is 
\begin{equation}
\label{eq: 1.5}
D_{\rm eff}\sim D \exp\left[-\frac{C(0)}{D^2}\right]
\ .
\end{equation}
An earlier work by De Gennes \cite{DeGennes1975} proposes a similar expression.
We discuss the origin of this result, and present a generalisation of it to non-Gaussian distributions, in section \ref{sec: 2}. 
When $D$ is small, this estimate for the diffusion coefficient depends upon 
rare events where the potential is unusually large, and it is 
very difficult to verify equation (\ref{eq: 1.5}) numerically.
In addition, numerical experiments show that the model exhibits sub-diffusive behaviour and 
it has been suggested that there is anomalous diffusion, in 
the sense that $\langle x^2\rangle \sim t^{\alpha}$, with $0<\alpha<1$ 
\cite{Khoury2011,Simon2013,Goychuk2017}. It is desirable to achieve 
an analytical understanding of the sub-diffusive behaviour which is observed 
in numerical simulations of (\ref{eq: 1.1}). 

We should mention that there are also exact results \cite{Sinai1982,Comtet1998,LeD+99} on a 
closely related model (motion in a quenched 
velocity field, which is \emph{not} the derivative of a potential with a well-defined 
probability distribution) showing that $\langle x^2\rangle\sim (\ln\, t)^{1/4}$. 
This \lq Sinai diffusion' process is fundamentally different, because the 
particle becomes trapped in successively deeper minima of the potential, 
from which it takes ever increasing time intervals to escape.

We will argue that, while equation (\ref{eq: 1.5}) and its generalisation to 
non-Gaussian distributions describes the long-time asymptote 
of the dispersion of particles, the diffusive behaviour only emerges at 
very long times. At intermediate times, the dynamics of typical realisations 
is not diffusive. We show that it is determined by the time taken to diffuse across the 
largest potential barrier which must be traversed to reach position $x$. The diffusion 
process is able to  traverse a barrier of height $\Delta V$ after a characteristic 
time $T\sim \exp(\Delta V/D)$ \cite{Kramers1940}, and as time increases the height of the barriers 
which can be breached, leading to \lq flooding' of the region beyond, increases. 
According to this picture, the dispersion distance $x$ is determined by a problem in extreme 
value statistics: how large must $x$ be before we reach a barrier of height $\Delta V\approx D\ln t$?
By considering the solution of this problem in extreme value statistics, we argue that the median 
$\bar T$ (with respect to different realisations of the potential $V(x)$) 
of the mean-first-passage-time (averaged over $\eta(t)$) satisfies
\begin{equation}
\label{eq: 1.6}
\ln\left(\frac{x}{\tilde x}\right)
\sim J(D\ln\bar T/2)
\end{equation}
where $\tilde x$ is a lengthscale which characterises the typical distance between 
extrema of the potential. In the case where the potential has a Gaussian distribution, 
this implies that the dispersion is sub-diffusive, satisfying 
\begin{equation}
\label{eq: 1.7}
\ln\left(\frac{x}{\tilde x}\right)\sim \frac{D^2(\ln \bar T)^2}{8C(0)}
\end{equation}
which is quite distinct from the usual anomalous diffusion behaviour, characterised
by power-laws such as $\langle x^2\rangle\sim t^\alpha$. 
After a sufficiently long time, the dynamics becomes diffusive, with a diffusion coefficient 
given by (\ref{eq: 1.5}).

Our arguments will depend upon making estimates of sums of the form 
\begin{equation}
\label{eq: 1.11}
S_N=\sum_{j=1}^N \exp(f_j/\epsilon)
\end{equation}
where $f_j$ are independent identically distributed (i.i.d.) random variables, and $\epsilon$ is a small 
parameter, which we identify with the diffusion coefficient $D$. We term the $S_N$ 
\lq extreme-weighted sums', because the largest values of 
$f_j$ make a dominant contribution to $S_N$ as $\epsilon\to 0$. In section \ref{sec: 2} we show how the 
mean-first-passage time is related to sums like (\ref{eq: 1.11}), and in section \ref{sec: 3} we analyse 
some of their statistics, which are used in section \ref{sec: 4} to justify our principle result, 
equation (\ref{eq: 1.6}). Section \ref{sec: 5} describes our numerical investigations, and 
section \ref{sec: 6} is a summary.

\section{The mean first passage time}
\label{sec: 2}

Our discussion of the dynamics of (\ref{eq: 1.1}) will focus on the mean first passage 
problem: what is the mean time $T(x)$ at which a particle released from  the origin 
reaches position $x$. First passage problems are discussed comprehensively in the book by Redner \cite{Red01}.  
The result that we require can be found in multiple sources: \cite{Lif+62} is the earliest 
reference that we are aware of and the key formula, equation (\ref{eq: 2.1}) below, was already 
applied to equation (\ref{eq: 1.1}) by Zwanzig \cite{Zwanzig1988}.

In this section we first quote the general formula for the mean first 
passage time $T(x)$, as a functional of the potential $V(x)$. 
If particles are released at $x_0=0$, the mean first passage time to reach 
position $x$ is given by
\begin{equation}
\label{eq: 2.1}
T(x)=\frac{1}{D}\int_0^x{\rm d}y\ \exp[V(y)/D]\int_0^y{\rm d}z\ \exp[-V(z)/D]
\end{equation}
where the averaging is with respect to realisations of the noise $\eta(t)$ in 
the equation of motion (\ref{eq: 1.1}), with $V(x)$ frozen, so that $T(x)$ is a functional of $V(x)$. 

We then (subsection \ref{sec: 2.1}) discuss 
the result obtained by Zwanzig \cite{Zwanzig1988} for the expectation value $\langle T(x)\rangle$ 
(averaged with respect to realisations of $V(x)$). 
Zwanzig gave the result for a potential with Gaussian fluctuations, which we extend to the case of a general 
form for the large-deviation entropy function (as defined by equation (\ref{eq: 1.4})). 
The result obtained by Zwanzig suggests that the dispersion is 
diffusive, with a diffusion coefficient $D_{\rm eff}$ which vanishes in a highly singular fashion as $D\to 0$. 
We shall argue that this result is a consequence of the expectation value of $T(x)$ being dominated 
by very rare large excursions of the potential $V(x)$, and that for typical realisations of the potential the 
dispersion is much more rapid than the value of $\langle T(x)\rangle$ suggests. This requires a more 
delicate analysis of the structure of the integrals in the expression for $T(x)$, equation (\ref{eq: 2.1}). 
In subsection \ref{sec: 2.2}, we discuss how these integrals may be approximated by sums, analogous to 
(\ref{eq: 1.11}), in the limit as $D\to 0$. 

\subsection{Expression for expectation value of mean first passage time}
\label{sec: 2.1}

We can make an additional average of (\ref{eq: 2.1}), with respect to different realisations of 
the potential, which leads to 
\begin{eqnarray}
\label{eq: 2.1.1}
\langle T(x)\rangle &=&\frac{1}{D}\int_0^x{\rm d}y\ \int_0^y{\rm d}z\ 
\bigg\langle \exp\left[\frac{V(y)-V(z)}{D}\right]\bigg\rangle
\nonumber \\
&\sim&\frac{x^2}{2D}\langle \exp(-V/D)\rangle\langle \exp(V/D)\rangle
\end{eqnarray}
where in the second line we consider the leading order behaviour as $x\to \infty$.
If the motion were simple diffusion, with $V=0$, equation (\ref{eq: 2.1}) would evaluate 
immediately to $\langle T\rangle=x^2/2D$, so that it is reasonable to identify $x^2/2\langle T\rangle$ as the effective 
diffusion coefficient. Hence, assuming that the PDF of $V(x)$ is symmetric between $V$ and $-V$, 
we have
\begin{equation}
\label{eq: 2.1.2}
D_{\rm eff}=\frac{D}{\left[\langle \exp(V/D)\rangle\right]^2}
\ .
\end{equation}
When $D$ is small, $\langle \exp(V/D)\rangle$ is dominated by the tail of the PDF of $V$, so that 
\begin{eqnarray}
\label{eq: 2.1.3}
\langle \exp(V/D)\rangle &=& \int_{-\infty}^\infty {\rm d}V P_V\exp(V/D)
\nonumber \\
&\sim &\int_{-\infty}^\infty {\rm d}V \exp[V/D-J(V)]
\nonumber \\
&\sim&\sqrt{\frac{2\pi}{J''(V^\ast)}} \exp[V^\ast/D-J(V^\ast)]
\end{eqnarray}
where $V^\ast$ is the stationary point of the exponent, satisfying 
\begin{equation}
\label{eq: 2.1.4}
DJ'(V^\ast)=1
\ .
\end{equation}
From this we obtain
\begin{equation}
\label{eq: 2.1.5}
D_{\rm eff}\sim \frac{DJ''(V^\ast)}{2\pi}\exp\left[2J(V^\ast)-\frac{2V^\ast}{D}\right]
\ . 
\end{equation}
In the Gaussian case, where 
\begin{equation}
\label{eq: 2.1.6}
J=\frac{V^2}{2C(0)}+\frac{1}{2}\ln(2\pi C(0))
\end{equation}
equation (\ref{eq: 2.1.5}) agrees with (\ref{eq: 1.5}).

\subsection{Summation approximations}
\label{sec: 2.2}

In order to understand the implications of equation (\ref{eq: 2.1}), we should consider 
the behaviour of the integral 
\begin{equation}
\label{eq: 2.2.1}
S(x)=\int_0^x{\rm d}y\ \exp[-V(y)/D]
\end{equation}
in the limit as $D\to 0$. 
When $D$ is small this quantity may be estimated from the minima of the potential: 
\begin{equation}
\label{eq: 2.2.2}
S(x)\sim \sum_{j=1}^N \sqrt{\frac{2\pi D}{V''^-_j}}\exp(-V_j^-/D)
\equiv \sum_{j-1}^N \exp[-\tilde V^-_j/D]
\end{equation}
where $V^-_j$ are the values of the $N$ minima between $0$ and $x$, occurring at positions $x^-_j$, 
and where we have defined 
\begin{equation}
\label{eq: 2.2.2a}
\tilde V^-_j=V_j^- - \frac{D}{2}\ln\left(\frac{2\pi D}{|V_j''^-|}\right) 
\end{equation}
Note that 
\begin{equation}
\label{eq: 2.2.3}
T(x)=\frac{1}{D}\int_0^x{\rm d}y\ \exp[V(y)/D]S(y)
\end{equation}
and consider how to estimate $T(x)$ in the limit as $D\to 0$. Note 
that $S(y)$ is determined by the values of the minima of $V(y)$ in the interval $[0,y]$, 
jumping by an amount $\exp[-\tilde V^-_j/D]$ at $x^-_j$. Similarly, if $V^+_j$ are local maxima of $V(x)$, occurring 
at positions $x^+_j$, then $T(x)$ jumps at local maxima. The evolution of $S(x)$ and $T(x)$ are therefore 
determined by a pair of coupled maps:
\begin{equation}
\label{eq: 2.2.4}
\begin{array}{ll}
S\rightarrow S'=S+\exp[-\tilde V_j^-/D]&\quad ({\rm at\ minima\ }x^-_j)\cr
\quad & \quad \cr
T\rightarrow T'=T+\frac{1}{D}\exp[\tilde V_j^+/D]S&\quad ({\rm at\ succeeding\ maximum\ }x^+_j)
\end{array}
\end{equation}
where we have defined again $\tilde{V}^+_j=V^+_j+\frac{D}{2}\ln\left(\frac{2\pi D}{|V''^+|}\right)$. 
These equations are difficult to analyse in the general case, but in the next section we discuss
an approach which can be used to treat the limit where $D$ is small.

\section{Statistics of extreme-weighted sums}
\label{sec: 3} 

We have seen that when $D$ is small the integrals defining the mean first passage 
time are approximated by sums over extrema of the potential, as described by equation 
(\ref{eq: 2.2.2}). Accordingly, we study properties of random sums of the form (\ref{eq: 1.11})
where $\epsilon$ is a small parameter and where the 
$f_j$ are drawn from a distribution for which the probability for $f_j$
being greater than $f$ is $Q(f)$. In the case where $f$ has a Gaussian distribution, 
(\ref{eq: 1.11}) is a sum of log-normal distributed random variables. There is some earlier 
literature on these sums which shows very little overlap with our results, see \cite{Rom+03} and 
references therein, also \cite{Pra+18}, which discusses a phase transition which arises in a limiting case.
We also consider sums of the form
\begin{equation}
\label{eq: 3.1}
T_N=\sum_{n=1}^N \exp(g_n/\epsilon)S_n
\end{equation}
where $g_j$ are drawn from the same i.i.d. distribution as the $f_j$. 
This is a model for the summation which approximates the integral $T(x)$ defined 
by equation (\ref{eq: 2.2.3}). When $\epsilon$ is sufficiently small, these sums are determined 
by the largest values of $f_j$ and $g_j$, and for this reason we shall refer to 
$S_N$ and $T_N$ as extreme-weighted sums.

We write the distribution function for $f$ in the form
\begin{equation}
\label{eq: 3.2}
Q(f)=\exp[-{\cal J}(f)]
\end{equation}
where ${\cal J}(f)$ is a large deviation rate function. 
We are interested in the asymptotic behaviour of statistics of the sums $S_N$ and $T_N$ for small $\epsilon$ and large $N$.
The sums vary wildly in magnitude and the mean is dominated by the tail of the distribution of $f$. 
Unless $N$ is sufficiently large, values of $f_j$ which determine the mean are unlikely to be sampled. 
This suggests that it will be useful to characterise the distribution of the $S_N$ by the median, 
rather than the mean. We denote the median of $X$ by $\bar X$ and its expectation by $\langle X\rangle$.

\subsection{Estimate of median of $S_N$}
\label{sec: 3.1}

The sum $S_N$ may be well approximated by its largest term, which is 
\begin{equation}
\label{eq: 3.1.1}
s_N=\exp(\hat f/\epsilon)
\end{equation}
where $\hat f$ is the largest of the $N$ realisations, $f_j$, with index $j=\hat j$. We write 
\begin{equation}
\label{eq: 3.1.2}
S_N\equiv \exp[\hat f/\epsilon]F\equiv s_N F
\end{equation}
where
\begin{equation}
\label{eq: 3.1.3}
F=1+\sum_{j=1\atop{j\ne \hat j}}^{N} \exp[-(\hat f-f_j)/\epsilon]
\ .
\end{equation}
If $F$ is close to unity, we can estimate $\bar S_N$ by $\bar s_N$. Let us first 
estimate $\bar s_N$ and return to consider $F$ later. Note that
\begin{equation}
\label{eq: 3.1.4}
\bar s_N=\exp(\bar {\hat f}/\epsilon)
\end{equation}
where $\bar {\hat f}$ is the median of the largest value of $N$ samples from the distribution of $f$.
This is determined by setting the probability for $N$ samples to be less than $f$ to be equal to one half:
\begin{equation}
\label{eq: 3.1.5}
\left[1-Q(\bar {\hat f})\right]^N=\frac{1}{2}
\ .
\end{equation}
When $N\gg 1$, this is determined by the tails of the distribution, where $Q(f)$ is approximated 
using (\ref{eq: 3.2}):
\begin{equation}
\label{eq: 3.1.6}
\exp\left[-N\exp[-{\cal J}(\bar {\hat f})]\right]=\frac{1}{2}
\end{equation}
so that $\bar {\hat f}$ satisfies
\begin{equation}
\label{eq: 3.1.7}
\ln N-\ln \ln 2={\cal J}(\bar {\hat f})
\ .
\end{equation}
An important special case is where the $f$ have a Gaussian distribution, so that
in the case where $\langle f\rangle=0$ and $\langle f^2\rangle=1$,
\begin{equation}
\label{eq: 3.1.8}
Q(f)=\frac{1}{\sqrt{2\pi}}\int_f^\infty {\rm d}x\ \exp(-x^2/2)
\sim \frac{1}{\sqrt{2\pi}f}\exp(-f^2/2)
\end{equation}
implying that
\begin{equation}
\label{eq: 3.1.9}
{\cal J}(f)=\frac{f^2}{2}+\ln\,f+\frac{\ln\,(2\pi)}{2}
\end{equation}
so that
\begin{equation}
\label{eq: 3.1.10}
\frac{\bar {\hat f}^2}{2}+\ln\, \bar {\hat f}=\ln N-\ln \ln 2-\frac{\ln\,(2\pi)}{2}
\ .
\end{equation}
In the limit where $N$ is extremely large, we can approximate $\bar {\hat f}$ by
\begin{equation}
\label{eq: 3.1.11}
\bar {\hat f}\sim\sqrt{2\ln\,N}
\end{equation}
and consequently the median of $\bar s_N$ is approximated by
\begin{equation}
\label{eq: 3.1.12}
\bar s_N\sim \exp\left(\frac{\sqrt{2\ln\,N}}{\epsilon}\right)
\ .
\end{equation}
Next consider how to estimate the quantity $F$ in equation (\ref{eq: 3.1.2}), when $\epsilon\ll 1$. 
When $N\gg 1$, either $F$ is close to unity or else it is the sum of a large number of terms 
which make a comparable contribution. The value of $F$ depends upon $\hat f$. The $f_j$ which 
contribute to $F$ are i.i.d. random variables, each with a PDF which is the same as that of the 
general $f_j$, except that there is an upper cutoff at $\hat f$: the adjustment of the normalisation 
due to the loss of the tail, $f>\hat f$, can be neglected. If the PDF of $f$ is
\begin{equation}
\label{eq: 3.1.13}
P_f=\exp[-J(f)]
\end{equation}
then the expectation value of $F$ is obtained as follows
\begin{eqnarray}
\label{eq: 3.1.14}
\!\!\!\!\!\!\!\!\!\!\!\!\!\!\!\exp(\hat f/\epsilon)[\langle F\rangle-1]&=&(N-1)\int_{-\infty}^{\hat f}{\rm d}f\ P_f\exp(f/\epsilon) 
\nonumber \\
\!\!\!\!\!\!\!\!\!\!\!\!\!\!\!&\sim & N\int_{-\infty}^{\hat f} {\rm d}f\ \exp[f/\epsilon-J(f)]
\nonumber \\
\!\!\!\!\!\!\!\!\!\!\!\!\!\!\!&\sim & \frac{N\exp[f^\ast /\epsilon-J(f^\ast)]}{\sqrt{J''(f^\ast)/2}}\int _{-\infty}^{\sqrt{J''(f^\ast)/2}(\hat f-f^\ast)}
{\rm d}y\ \exp(-y^2)
\end{eqnarray}
where $f^\ast$ satisfies 
\begin{equation}
\label{eq: 3.1.14a}
\epsilon J'(f^\ast)=1
\ .
\end{equation}
Noting that 
\begin{equation}
\label{eq: 3.1.15}
\langle S_N\rangle=N\langle \exp(f/\epsilon)\rangle\sim \sqrt{2\pi }\frac{N\exp[f^\ast/\epsilon-J(f^\ast)]}
{\sqrt{J''(f^\ast)}}
\end{equation}
we have 
\begin{equation}
\label{eq: 3.1.16}
\exp(\hat f/\epsilon)[\langle F\rangle-1]\sim \langle S_N\rangle \frac{1}{2}
\left[1+{\rm erf}\left(\sqrt{\frac{J''(f^\ast)}{2}}(\hat f-f^\ast)\right)\right]
\ .
\end{equation}
Hence, we obtain a rather simple approximation for $S_N$, depending upon the extreme 
value $\hat f$ of the sample of $N$ realisations of the $f_j$:
\begin{equation}
\label{eq: 3.1.17}
S_N\sim \exp(\hat f/\epsilon)+\langle S_N\rangle \frac{1}{2}
\left[1+{\rm erf}\left(\sqrt{\frac{J''(f^\ast)}{2}}(\hat f-f^\ast)\right)\right]
\ .
\end{equation}
The median of $S_N$ is therefore approximated by
\begin{equation}
\label{eq: 3.1.18}
\bar S_N\sim \exp(\bar {\hat f}/\epsilon)+\langle S_N\rangle \frac{1}{2}
\left[1+{\rm erf}\left(\sqrt{\frac{J''(f^\ast)}{2}}(\bar {\hat f}-f^\ast)\right)\right]
\end{equation}
where $\bar {\hat f}$ is the solution of equation (\ref{eq: 3.1.10}).

\subsection{Interpretation and generalisation to $T_N$}
\label{sec: 3.2}

We shall see that equation (\ref{eq: 3.1.18}) gives a quite precise approximation for the median $\bar S_N$, but 
it is not immediately clear when either of the two terms is dominant. In order to clarify the structure of 
equation (\ref{eq: 3.1.18}), we consider an approximate form of the equation determining $\bar{\hat f}$, 
and transform to logarithmic variables. As well as leading to a transparent understanding of equation 
(\ref{eq: 3.1.18}), this facilitates making an estimate for $\bar T_N$ in the limit where $\epsilon \ll 1$ 
and $N\gg 1$. We define
\begin{equation}
\label{eq: 3.2.1}
\eta=\ln\,N
\ ,\ \ \ 
\sigma=\ln\,\bar S_N
\ ,\ \ \ 
\tau=\ln\, \bar T_N 
\ .
\end{equation}
Note that $\eta$ and $\tau$ are logarithmic measures of, respectively, distance 
and time, so that a plot of $\eta$ versus $\tau$ gives information about the dispersion 
due to the dynamics. 

Let us consider the limit where the first term in (\ref{eq: 3.1.18}), $\exp[\bar{\hat f}/\epsilon]$, 
is dominant. Note that the condition (\ref{eq: 3.1.10}), determining the extreme value of a set of 
$N$ samples, can be approximated by the requirement that the PDF of $f$ is approximately 
equal to $1/N$. That is $1\sim N\exp[-J(\bar{\hat f})]$. For the purposes of considering 
the $N\to \infty$ and $\epsilon\to 0$ limit, we can therefore approximate $\bar{\hat f}$ by a solution 
of the equation
\begin{equation}
\label{eq: 3.2.2}
J(\bar{\hat f})=\eta
\ .
\end{equation}
If the second term in equation (\ref{eq: 3.1.18}) is negligible, as might be expected 
when $\bar {\hat f}-f^\ast\ll 1$, equations (\ref{eq: 3.2.1}) and (\ref{eq: 3.2.2}) then yield
a simple implicit equation for $\sigma$: 
\begin{equation}
\label{eq: 3.2.3}
\eta = J(\epsilon \sigma)
\ .
\end{equation}
If  $\bar {\hat f}-f^\ast\gg 1$, and if the second term in (\ref{eq: 3.1.18}) is dominant, then 
$\bar S_N\sim \langle S_N\rangle =N\langle \exp(f/\epsilon)\rangle$, and using the Laplace 
principle we find
\begin{equation}
\label{eq: 3.2.4}
\sigma=\eta+\frac{f^\ast}{\epsilon}-J(f^\ast)
\ .
\end{equation}
Note the (\ref{eq: 3.2.4}) indicates that $\frac{{\rm d}\sigma}{{\rm d}\eta}=1$. Let us compare this with 
the value of $\frac{{\rm d}\sigma}{{\rm d}\eta}$ obtained from (\ref{eq: 3.2.3}), which predicts 
$\frac{{\rm d}\eta}{{\rm d}\sigma}=\epsilon J'(f)$. The approximation (\ref{eq: 3.2.3}) therefore becomes 
sub-dominant when $1=\epsilon J'(f)$, which is precisely the equation for $f^\ast$,  equation 
(\ref{eq: 3.1.14a}). If we define $\eta^\ast$ and $\sigma^\ast$ by writing
\begin{equation}
\label{eq: 3.2.5}
\sigma^\ast=\frac{f^\ast}{\epsilon}
\ ,\ \ \ 
\eta^\ast=J(\epsilon\sigma^\ast)
\end{equation}
then assembling these results and definitions, the relationship between $\eta$ and $\sigma$ can be 
summarised in the following equation
\begin{equation}
\label{eq: 3.2.6}
\eta=\biggl\{
\begin{array}{cc}J(\epsilon \sigma)& 0<\eta<\eta^\ast\\
\eta=\sigma-\sigma^\ast+\eta^\ast & \eta\ge \eta^\ast
\end{array}
\ .
\end{equation}
Note that $\eta(\sigma)$ and its first derivative are continuous functions.
In the foregoing we defined $\bar x$ as the median of $x$, but it should be noted that our arguments 
will lead to equations (\ref{eq: 3.2.4}) and (\ref{eq: 3.2.5}) as $N\to \infty$ if $\bar S_N$ denotes 
any fixed percentile of $S_N$.

Thus far we have considered the behaviour of $\eta$ as a function of $\sigma$ rather than of $\tau$, 
but it is the function $\eta(\tau)$ which describes the dynamics of the dispersion.
Consider the form of the sum $T_N$ defined in equation (\ref{eq: 3.2}). 
When $\sigma<\sigma^\ast$, the value of $S_n$ is almost always determined by 
$\hat f$ the largest value of $f_j$, and similarly, one the factors $\exp(g_j/\epsilon)$
corresponding to $\hat g$, the largest of the $g_j$, will 
predominate over the others. In one half of realisations, those where $\hat k(N)>\hat j(N)$, 
the largest value of $f_j$ contributes to the sum which is multiplied by $\exp(\hat g/\epsilon)$, 
and we have $T_N\sim \exp(\hat g/\epsilon)\exp(\hat f/\epsilon)$. In cases where $\hat j>\hat k$, 
$T_N$ is expected to be small in comparison to this estimate. 
Noting that $\exp(\hat f/\epsilon)$ and $\exp(\hat g/\epsilon)$ 
are independent and both have probability one half to exceed $\exp(\bar {\hat f}/\epsilon)$ 
and $\exp(\bar {\hat g}/\epsilon)$ respectively, there are one quarter of realisations where 
$\exp[(\hat f^+\hat g)/\epsilon]$ exceeds $\exp[(\bar {\hat f}+\bar {\hat g})/\epsilon]$ and in 
half of these realisations $T_N\ll \exp[(\hat f+\hat g)/\epsilon]$.
If we now use the overbar to represent the upper octile of the distribution of $T_N$, 
rather than the median, we have
\begin{equation}
\label{eq: 3.2.7}
\bar T_N\sim \exp(\bar {\hat g}/\epsilon)\exp(\bar {\hat f}/\epsilon)
\ .
\end{equation}
Using the assumption that the $f_j$ and $g_j$ have the same PDF, we can conclude that 
$\bar T_N\sim \bar S_N^2$ and hence that $\tau =2\sigma$. The equation describing the dispersion 
as a function of time is therefore 
\begin{equation}
\label{eq: 3.2.8}
\eta=J(\epsilon \tau/2)
\ ,\ \ \ \tau<\tau^\ast
\end{equation}
where $\tau^\ast$ is determined by the condition that ${\rm d}\eta/{\rm d}\tau=\frac{1}{2}$ when $\tau=\tau^\ast$.
When $\tau>\tau^\ast$, we have $\bar T_N\sim \langle T\rangle =N\langle \exp(f/\epsilon)\rangle^2$, 
implying that 
\begin{equation}
\label{eq: 3.2.9}
\eta=J(\epsilon\tau^\ast/2)+\frac{\tau-\tau^\ast}{2}
\ ,\ \ \ 
\tau>\tau^\ast
\ .
\end{equation}
Equations (\ref{eq: 3.2.8}) and (\ref{eq: 3.2.9}) are a description of the logarithm of the typical dispersion $\eta$ 
as a function of the logarithm of the time, $\tau$. Usually the function $J(V)$ has a quadratic 
behaviour for small values of $V$, so that the initial dispersion, described by (\ref{eq: 3.2.8}), 
is sub-diffusive. The factor of one half in (\ref{eq: 3.2.9}) indicates that the long-time limit is diffusive. 
Writing $\eta^2\sim \tau+\ln D_{\rm eff}$, we see that the effective diffusion coefficient is 
\begin{equation}
\label{eq: 3.2.10}
D_{\rm eff}\sim \exp\left[2J(\epsilon\tau^\ast/2)-\tau^\ast \right] 
\end{equation}
which is consistent with (\ref{eq: 2.1.5}).

\section{Flooding dynamics model for dispersion}
\label{sec: 4}

In section \ref{sec: 2}, we showed that the integrals which are used to compute the mean-first-passage time
may be approximated by sums when $D$ is small. In section \ref{sec: 3}, we considered the statistics 
of these sums, $S_N$ and $T_N$, defined by equations (\ref{eq: 1.11}) and (\ref{eq: 3.1}) respectively. 
In terms of the calculation discussed in section \ref{sec: 2}, our estimate of $\bar T_N$ corresponds, 
for $N<N^\ast$, to the value of $\bar T(x)$ being determined by the difference between the lowest minimum 
of the potential and its highest maximum, provided the minimum occurs before the maximum.
We can therefore think of $\bar T(x)$ being determined by a \lq flooding' model, according to which 
the probability density for locating the particle occupies a region which is constrained by a potential 
barrier which can trap a particle for time $\bar T$. As $\bar T$ increases,  higher 
barriers are required.

In terms of the original problem, discussed in section \ref{sec: 2}, $N$ is the number of extrema 
of the potential before we reach position $x$. The arguments of section \ref{sec: 3} imply that the 
upper octile of the mean-first-passage time, $\bar T(x)$, satisfies an 
equation similar to (\ref{eq: 3.2.7}). We define logarithmic variables 
\begin{equation}
\label{eq: 4.1}
\eta=\ln\left(\frac{x}{\tilde x}\right)
\ ,\ \ \ 
\tau =\ln \bar T(x)
\end{equation}
where $\tilde x$ is the mean separation of minima of $V(x)$.
In terms of these logarithmic variables, the dispersion is described by
\begin{equation}
\label{eq: 4.2}
\eta=J\left(D\tau /2\right)
\ .
\end{equation}
which is valid up to $\tau^\ast$, which is defined by the condition 
\begin{equation}
\label{eq: 4.3}
\frac{{\rm d}\eta}{{\rm d}\tau}\bigg\vert_{\tau^\ast}=\frac{1}{2}
\ .
\end{equation}
Equation (\ref{eq: 4.2}) is our principal result. It applies to any percentile of the distribution which remains fixed 
when we take the limits $N\to \infty$ and $\epsilon\to 0$. 
When $\tau$ is large compared to $\tau^\ast$, equation (\ref{eq: 4.2}) is replaced by a linear 
relation, with an effective diffusion coefficient $D_{\rm eff}$
\begin{equation}
\label{eq: 4.4}
2\eta\sim \tau+\ln(D_{\rm eff}/D)
\ ,\ \ \ 
D_{\rm eff}\sim D\exp\left[2J(D\tau^\ast/2)-\tau^\ast \right]
\ .
\end{equation}

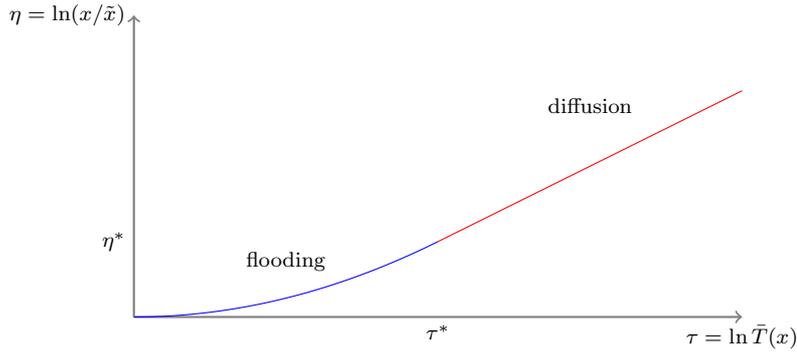
\begin{figure}[t!]
\centering 
\label{fig: 0}
\begin{tikzpicture}
\draw [thick, gray, ->] (0,0) -- (0,4)  
        node [left, black] {$\eta=\ln(x/\tilde x)$};  
\draw [thick, gray, ->] (0,0) -- (8,0)  
        node [below, black] {$\tau=\ln \bar T(x)$}; 
\node [left] at (0,1) {$\eta^\ast$};  
\node [below] at (4,0) {$\tau^\ast$};  
\node [above] at (2,0.5) {flooding};  
\node [below] at (6,3) {diffusion};  
\draw[scale=1.0, domain=0:4, smooth, variable=\x, blue] plot ({\x}, {\x*\x/16});
\draw[scale=1.0, domain=4:8, smooth, variable=\x, red]  plot ({\x}, {-1.0+0.5*\x});   
\end{tikzpicture}
\caption{The dynamics of a typical realisation, characterised by the median $\bar T$ of the 
mean-first-passage time, shows a crossover from sub-diffusive \lq flooding' dynamics to slow diffusion.}
\end{figure}

An important example is the case where $V$ has a Gaussian distribution, so that $J\sim V^2/2C(0)$. In terms
of the diffusion coefficient $D$, equations (\ref{eq: 4.2})-(\ref{eq: 4.4}) give
\begin{eqnarray}
\label{eq: 4.5}
\eta \sim \frac{D^2 \tau^2}{8C(0)} &\quad & \tau<\tau^\ast=\frac{2C(0)}{D^2}
\nonumber \\
\eta \sim \frac {\tau}{2}-\frac{C(0)}{2D^2} &\quad & \tau>\tau^\ast
\end{eqnarray} 
and using (\ref{eq: 4.4}) we find $D_{\rm eff}\sim D\exp[-C(0)/D^2]$, in agreement with (\ref{eq: 1.5}).
A sketch of the dependence of $\eta$ upon $\tau$ for the Gaussian case is shown in Fig.~\ref{fig: 0}. 

\section{Numerical studies}
\label{sec: 5}

We performed a variety of numerical investigations, using Gaussian distributed random 
variables $f_j$ to test the theory of extreme-weighted sums, and a Gaussian random function $V(x)$ 
to test the analysis of continuous potentials. In both cases the Gaussian variables had zero mean
and unit variance. In the case of the random potential, we also used a Gaussian for the correlation 
function, with a correlation length of order unity:
\begin{equation}
\label{eq: 5.1}
\langle V(x)V(x')\rangle=\exp[-(x-x')^2/2]
\ .
\end{equation}

\subsection{Discrete sums}
\label{sec: 5.1}

We characterised the statistics of the discrete sum (\ref{eq: 1.11}) by making 
a careful estimate of its median, equation (\ref{eq: 3.1.18}). 
In order to evaluate equation (\ref{eq: 3.1.18}), we need a solution of the implicit equation (\ref{eq: 3.1.10}), 
which determines $\bar {\hat f}$. By substituting (\ref{eq: 3.1.11}) into (\ref{eq: 3.1.10}), we find 
\begin{equation}
\label{eq: 5.2}
\bar {\hat f}\approx \sqrt{2\ln N-\ln\,\ln\,N-\ln\,(2\pi)-2\ln\,\ln\,2}
\ .
\end{equation}
The expression for the median approaches that for the mean 
\begin{equation}
\label{eq: 5.3}
S_N\sim N\exp\left(\frac{1}{2\epsilon^2}\right)
\end{equation}
at large values of $N$ when $\bar{\hat f}$ exceeds $f^\ast=1/\epsilon$.

For very large $N$ and very small $\epsilon$, the medians of $S_N$ and $T_N$ 
are estimated by simplified expressions, relating $\sigma=\ln \bar S_N$ 
and $\tau=\ln \bar T_N$ to $\eta=\ln N$. In the Gaussian case, these equations 
(\ref{eq: 3.2.6}), (\ref{eq: 3.2.8}) and (\ref{eq: 3.2.9}) give
\begin{equation}
\label{eq: 5.4}
\sigma=\biggl\{
\begin{array}{cc}
\frac{\sqrt{2\eta}}{\epsilon}& 0<\eta<\eta^\ast\\
\eta-\eta^\ast+\sigma^\ast & \eta\ge \eta^\ast
\end{array}
\end{equation}
and
\begin{equation}
\label{eq: 5.5}
\tau=\biggl\{
\begin{array}{cc}\frac{\sqrt{8\eta}}{\epsilon}& 0<\eta<\eta^\ast\\
2(\eta-\eta^\ast)+\tau^\ast&\eta>\eta^\ast
\end{array}
\end{equation}
where
\begin{equation}
\label{eq: 5.6}
\eta^\ast=\frac{1}{2\epsilon^2}
\ ,\ \ \ 
\sigma^\ast=\frac{1}{\epsilon^2}
\ ,\ \ \ 
\tau^\ast=\frac{2}{\epsilon^2}
\ .
\end{equation}
These equations imply that, in the limit as $\epsilon\to 0$, if we plot $y=\sigma/\eta^\ast$ as 
function of $x=\eta/\eta^\ast$, the numerical data for $\bar S_N$ should collapse onto the 
function 
\begin{equation}
\label{eq: 5.7}
y=f(x)=\biggl\{
\begin{array}{cc}2\sqrt{x}& 0<x<1\\
x+1& x>1
\end{array}
\ .
\end{equation}
Similarly, $y'=\tau/\eta^\ast$ plotted as a function of $x=\eta/\eta^\ast$ should collapse to $y'=2f(x)$.

We computed $M\in\{10,100,1000\}$ realisations of the sums $S_N$ and $T_N$, for 
$\epsilon\in\{1/3,1/4,1/6,1/8\}$ and $N\le 10^5$ 
(except for $M=1000$, in which case $N\le 5\times 10^4$). We evaluated the sample 
average $\langle S_N\rangle_M$, the sample median, $\bar S_N\vert_{M,2}$ and the sample 
upper octile $\bar S_N\vert_{M,8}$. We also computed the same statistics for the $T_N$.

\begin{figure}[h]
\centering
\begin{tabular}{c}
\epsfig{file=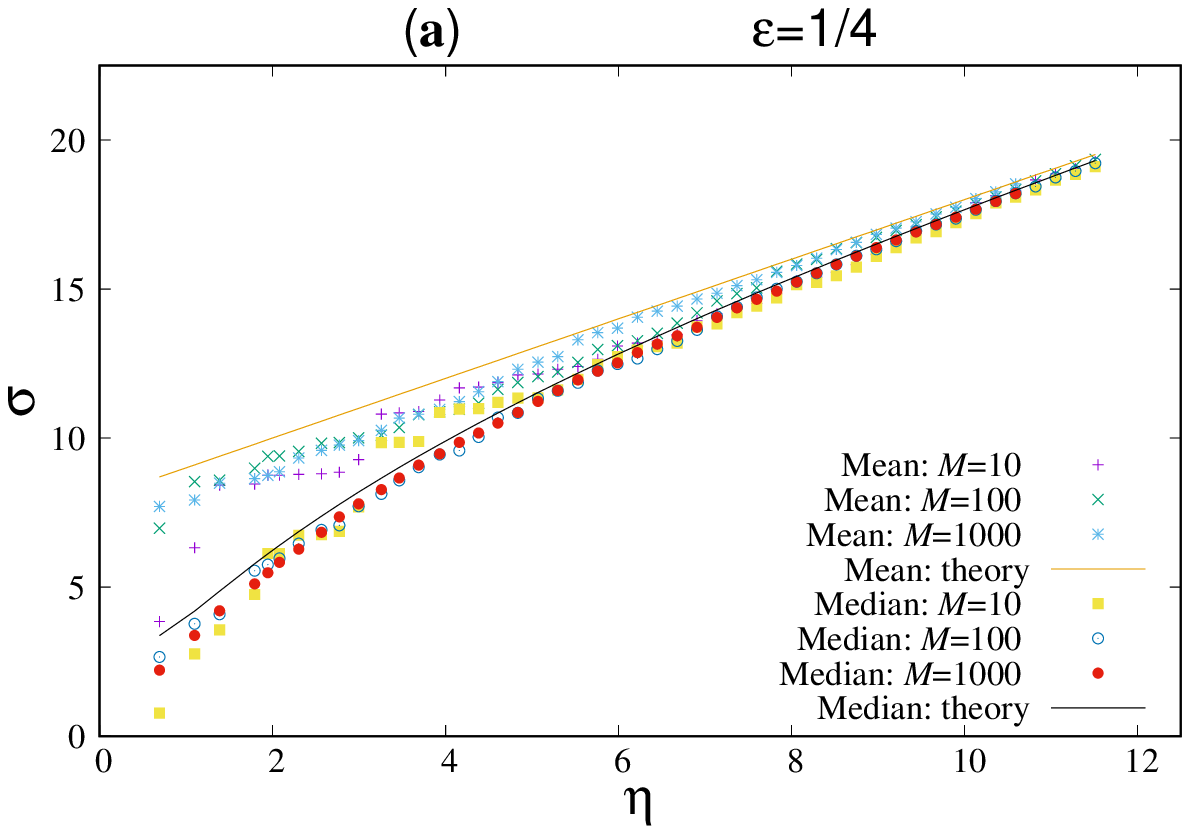,width=0.6\linewidth,clip=} \\
\epsfig{file=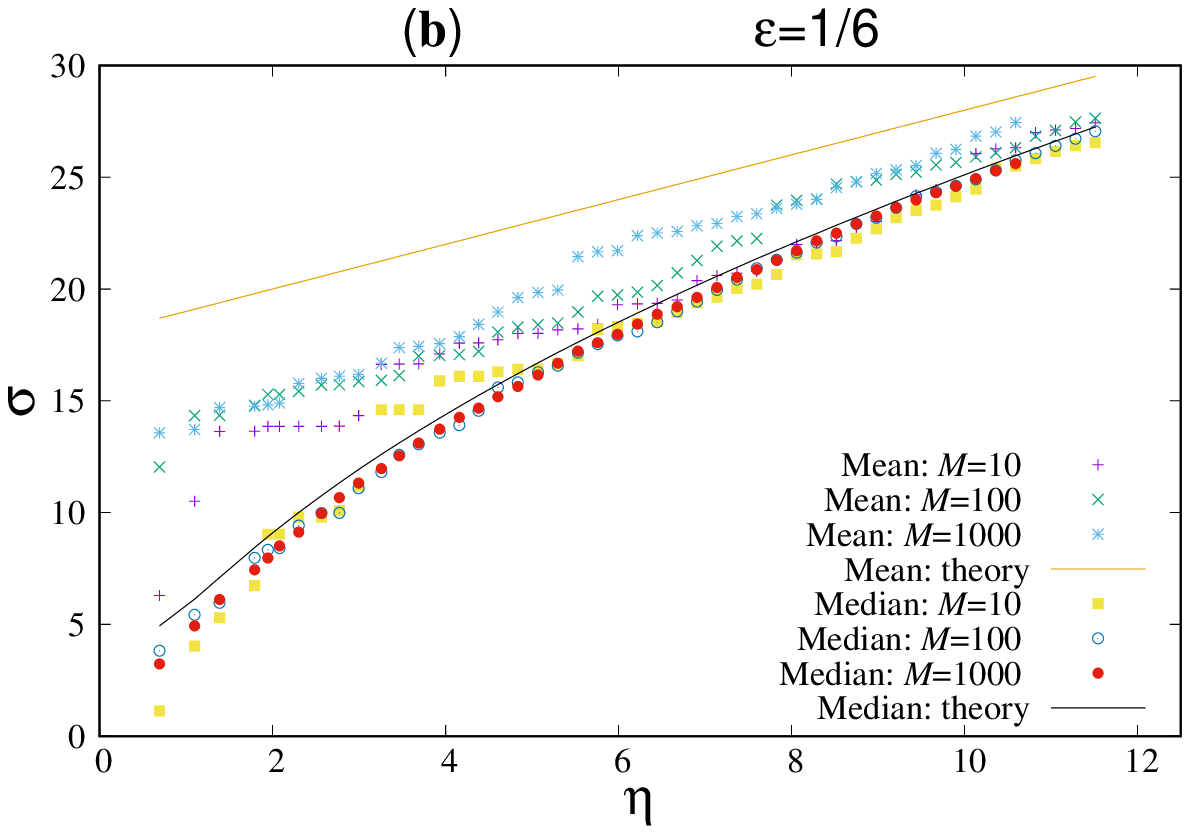,width=0.6\linewidth,clip=}
\end{tabular}
\caption{
\label{fig: 1}
Plot of $\ln \bar S_N$ and $\ln \langle S_N\rangle$, as a function of $\eta= \ln N$, 
for $\epsilon=1/4$ ({\bf a}) and $\epsilon=1/6$ ({\bf b}).
}
\end{figure}

Figure \ref{fig: 1} plots $\ln \bar S_N$, and $\ln \langle S_N\rangle$ as a function of $\eta=\ln N$, 
for different sample sizes, for $\epsilon=1/4$ ({\bf a}) and $\epsilon=1/6$ ({\bf b}). We compare with the 
theoretical prediction, obtained from (\ref{eq: 3.1.18}) and (\ref{eq: 5.2}) (for the median) and (\ref{eq: 5.3})
(for the mean). The agreement with the theory for the median is excellent. Note that the convergence of the mean 
value for different sample sizes is very poor when $\eta<\eta*=1/2\epsilon^2$ (this  
is especially apparent for smaller values of $\epsilon$).

\begin{figure}[h]
\centering
\begin{tabular}{cc}
\epsfig{file=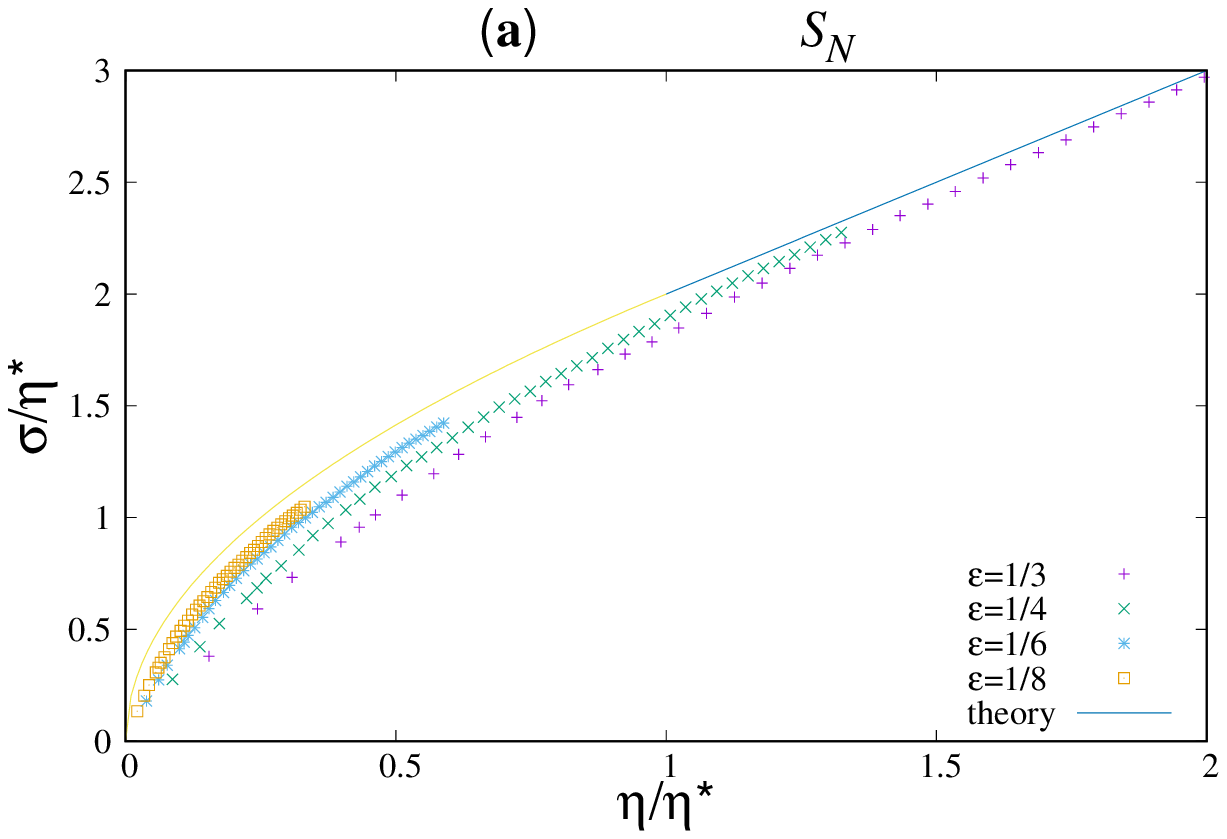,width=0.5\linewidth,clip=} &
\epsfig{file=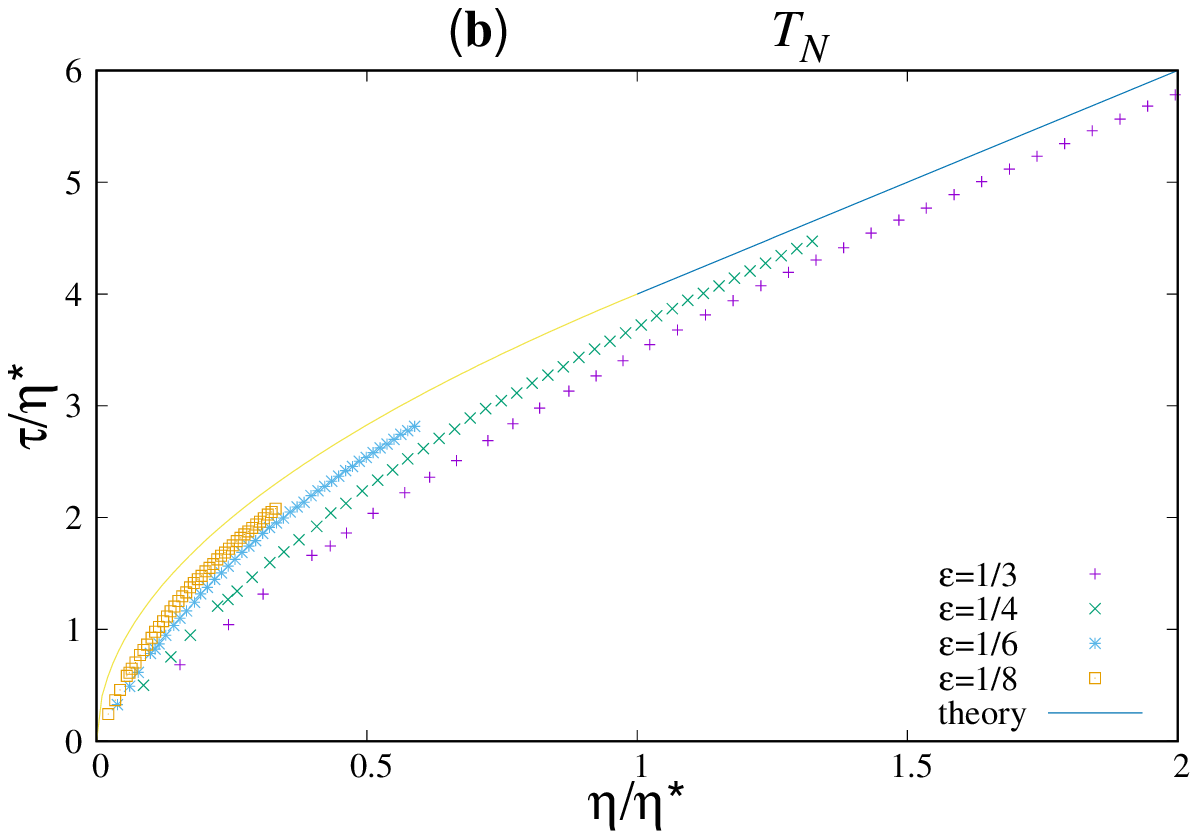,width=0.5\linewidth,clip=}\\
\epsfig{file=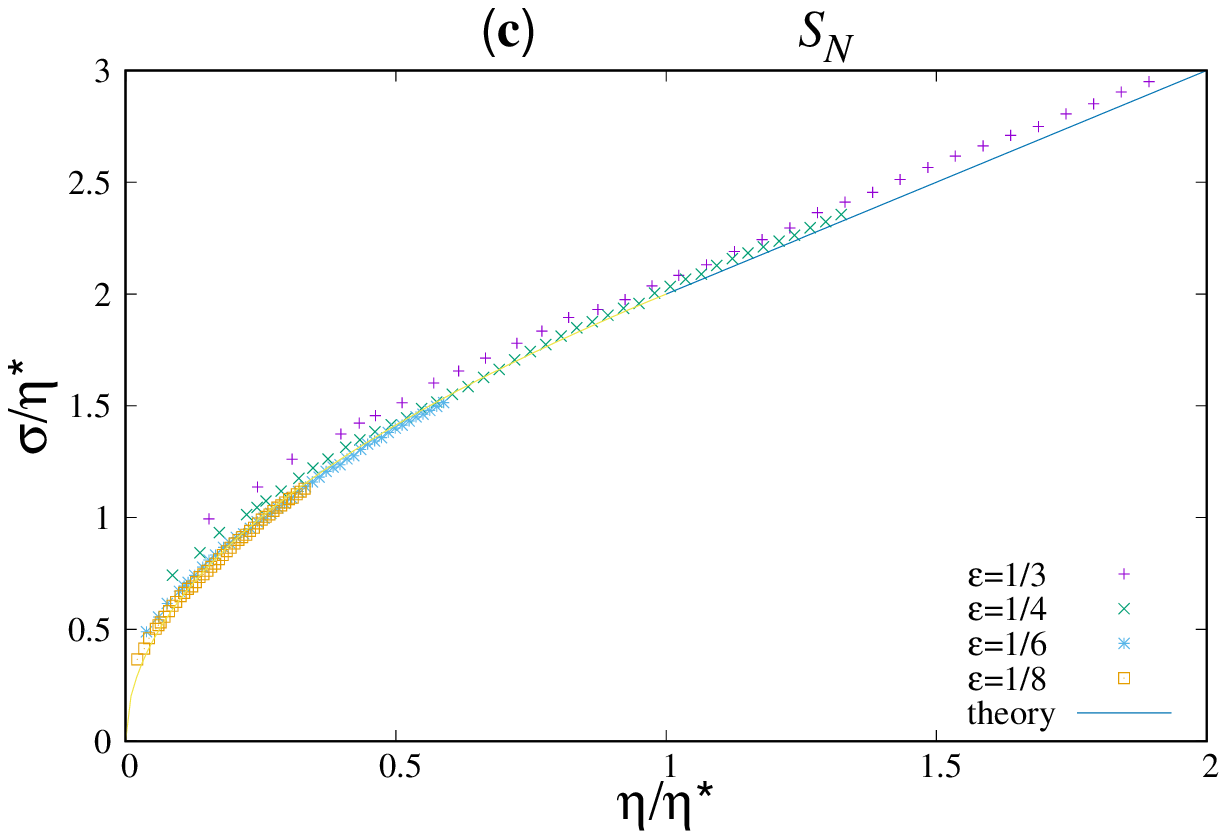,width=0.5\linewidth,clip=} &
\epsfig{file=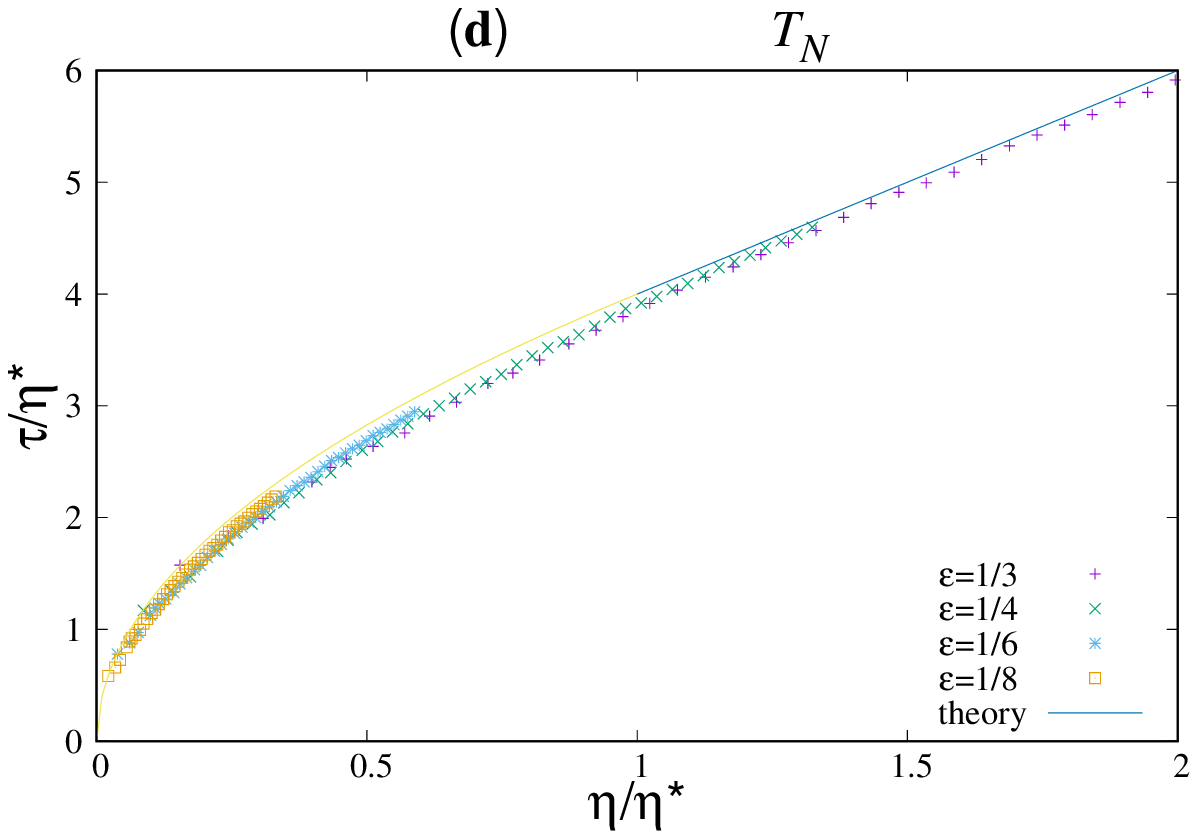,width=0.5\linewidth,clip=}
\end{tabular}
\caption{
\label{fig: 2}
Plot of $\sigma/\eta^\ast$ ({\bf a}) and $y'=\tau/\tau^\ast$ ({\bf b}) based upon median values, 
as a function of $\eta/\eta^\ast$, compared with the theoretical prediction for the $\epsilon\to 0$ limit, 
equation (\ref{eq: 5.7}). In ({\bf c}) and ({\bf d}) we show similar plots for the upper octile data.
}
\end{figure}

In figure \ref{fig: 2}, we plot $y=\sigma/\sigma^\ast$ ({\bf a}) and $y'=\tau/\tau^\ast$ ({\bf b}) as a function of 
$x=\eta/\eta^\ast$, for all of the values of $\epsilon$ in our data set, using the largest sample size ($M=1000$)
in each case, comparing with the theoretical scaling function (\ref{eq: 5.7}). We see convergence towards 
the function (\ref{eq: 5.7}) as $\epsilon\to 0$. In panels ({\bf c}) and ({\bf d}), we make the same comparison using the 
upper octile rather than the median.

\subsection{Continuous potentials}
\label{sec: 5.2}

\begin{figure}[h]
\includegraphics[width=1.0\linewidth]{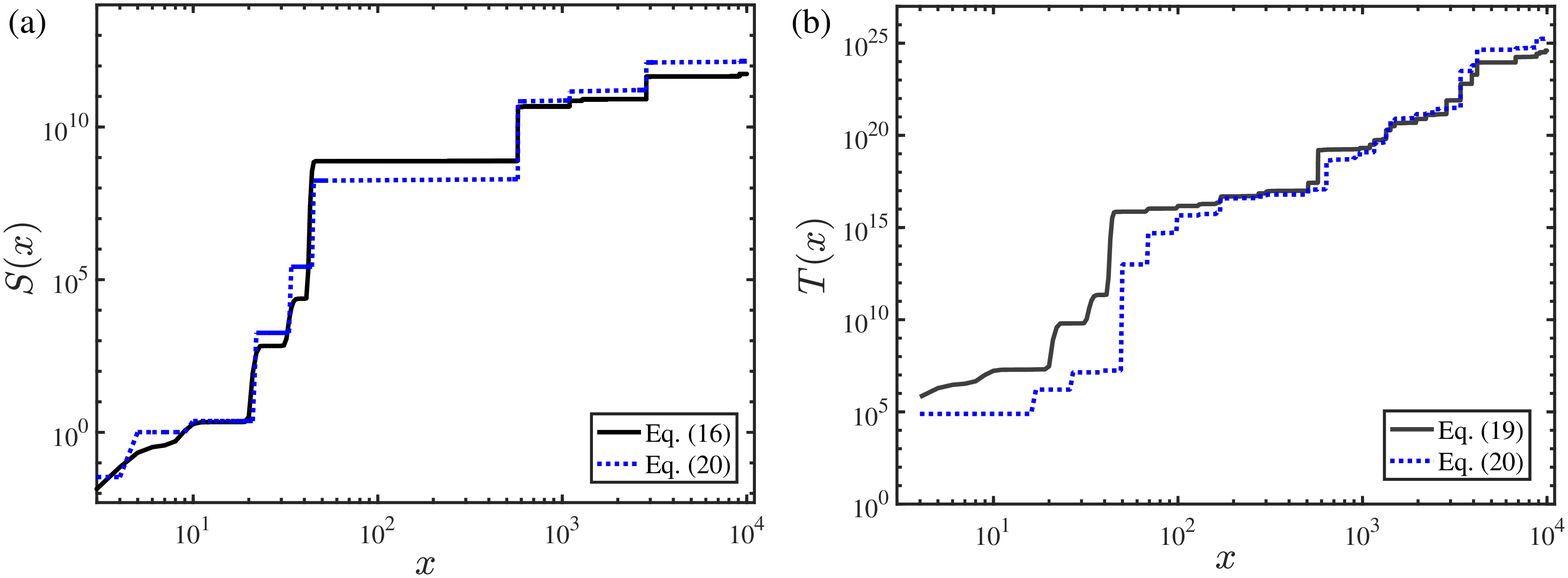}
\caption{
\label{fig: 3}
Plots of $S(x)$ ({\bf a}) and $T(x)$ ({\bf b}), for one realisation of the smooth random potential, with $D=1/8$. 
The numerically evaluated integrals (solid curves) are in good agreement with approximations based
on sums over maxima and minima (dashed lines).
}
\end{figure}

In the case of a continuous potential, we require the mean separation of maxima or minima, $\tilde x$, in order 
to make a comparison with theory. The density ${\cal D}$ of zeros of $V'(x)$ may be determined 
by the approach developed by Kac \cite{Kac1943} and Rice \cite{Rice1945}. If ${\cal P}(V,V',V'')$ is the joint 
PDF of $V(x)$ and its first two derivatives, evaluated at the same point, we find that
\begin{equation}
\label{eq: 5.8}
{\cal D}=\int_{-\infty}^\infty{\rm d}V\int_{-\infty}^\infty{\rm d}V''\ {\cal P}(V,0,V'')|V''|
\ .
\end{equation}
By noting that the vector $(V,V',V'')$ has a multivariate Gaussian distribution, and expressing 
${\cal P}(V,V',V'')$ in terms of the correlation function of the elements of this vector, we obtain
${\cal D}$ and hence the separation of minima $\tilde x$ for the potential satisfying (\ref{eq: 5.1}):
\begin{equation}
\label{eq: 5.9}
\frac{2}{\tilde x}\equiv {\cal D}=\frac{\sqrt{3}}{2\pi}
\ .
\end{equation}
First we investigated whether the mean-first passage time can be accurately represented 
by sums over maxima and minima of the potential. In figure \ref{fig: 3}, we compare the 
numerical evaluation of the integrals $S(x)$ ({\bf a}) and $T(x)$ ({\bf b}), 
given by Eq.~(\ref{eq: 2.2.1}) and Eq.~(\ref{eq: 2.2.3}), respectively, with the 
approximations which estimate the integrals using maxima and minima, Eq.~(\ref{eq: 2.2.4}).

\begin{figure}[h]
\centering
\begin{tabular}{cc}
\epsfig{file=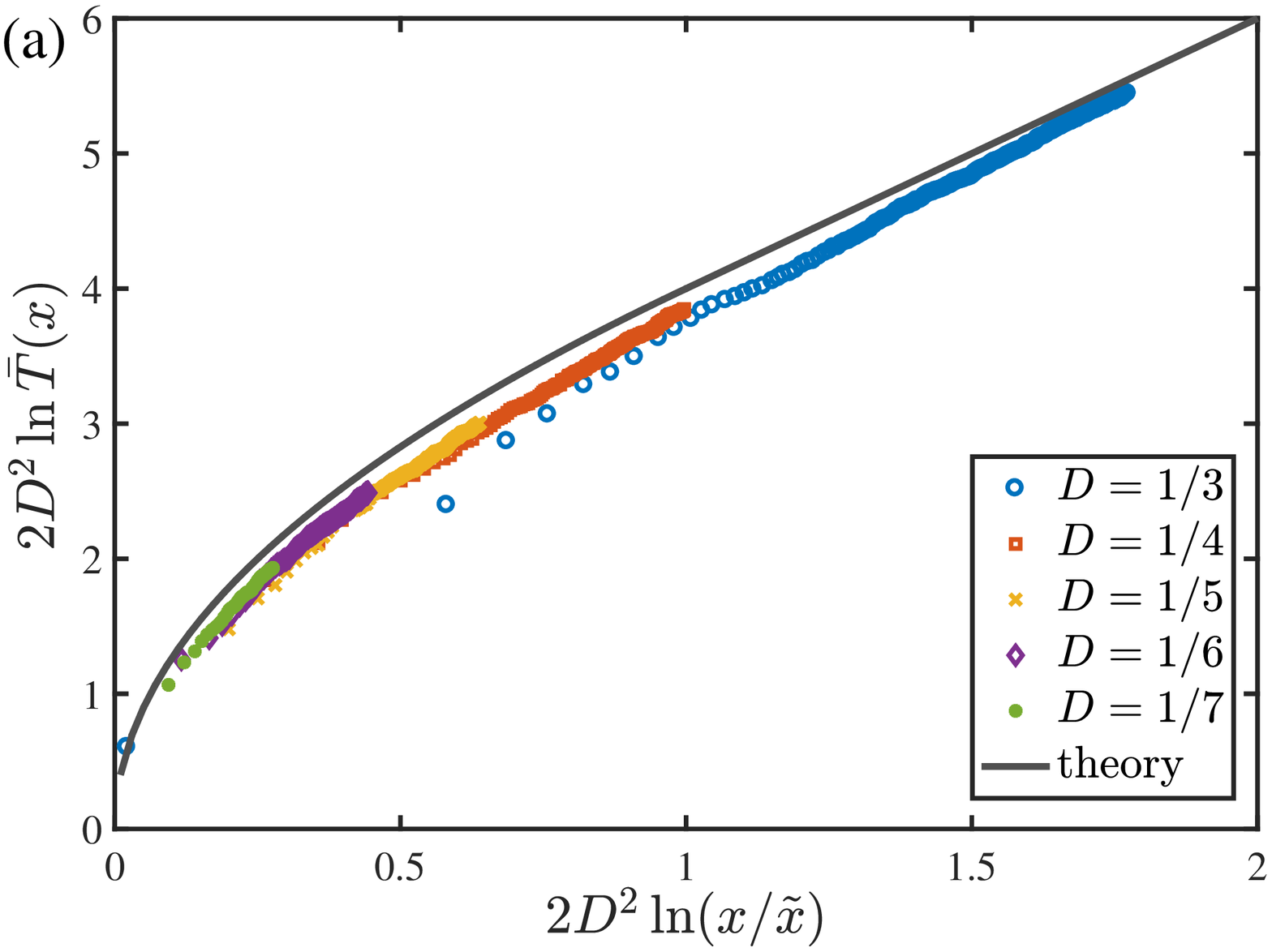,width=0.48\linewidth,clip=} &
\epsfig{file=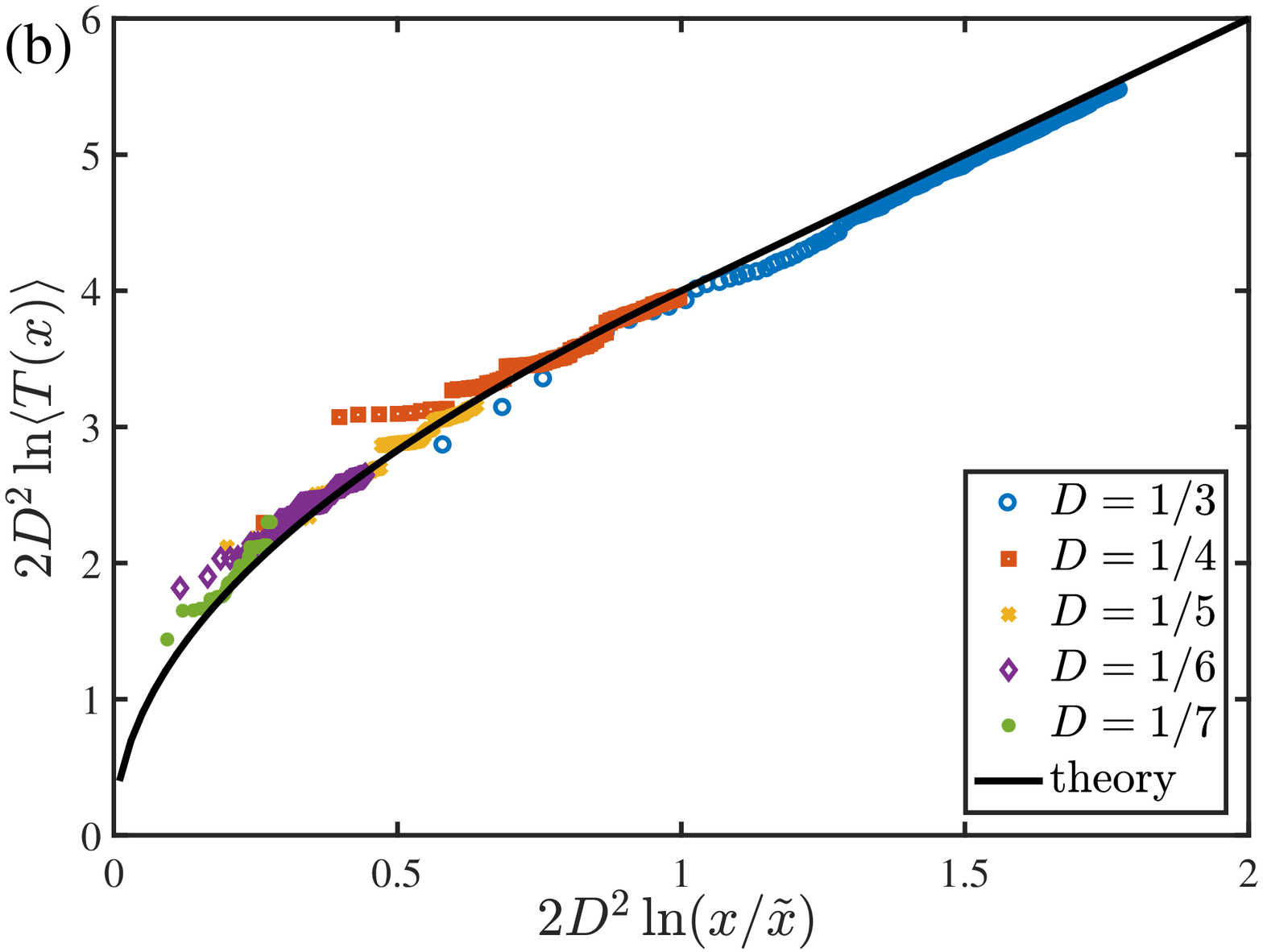,width=0.48\linewidth,clip=}
\end{tabular}
\caption{
\label{fig: 4}
Numerical results on the median $\bar T(x)$ of the mean 
first-passage time to reach $x$, for different values of the diffusion coefficient $D$. We plot  
logarithmic variables $Y_{\rm med}=D^2\ln \bar T(x)/2$ ({\bf a}) and 
$Y_{\rm av}=D^2\ln \langle T(x)\rangle$ ({\bf b}) as a function of $X=2D^2\ln (x/\tilde x)$ 
(where $\tilde x$ is the mean distance between maxima of $V(x)$). The numerical results converge 
towards the theoretically predicted scaling function, equation (\ref{eq: 5.7}), as $D\to 0$.
}
\end{figure}

We evaluated the median $\bar T(x)$ and mean $\langle T(x)\rangle$ of the 
mean first passage time $T(x)$ for $1000$ realisations of the potential $V(x)$, 
up to $x_{\rm max}=10^4$, for $D\in\{1/3,1/4,1/5,1/6,1/7\}$. According to the discussion 
in section \ref{sec: 4}, we expect that $\tau=\ln \bar T(x)$ and $\eta=\ln(x/\tilde x)$ 
are related by $\eta=J(D\tau/2)=D^2\tau^2/8$, up to a maximum value of $\eta$, 
given by $\eta^\ast=1/2D^2$. In figure \ref{fig: 4} we plot $Y_{\rm med}=2D^2\ln \bar T(x)$ and 
$Y_{\rm av}=2D^2\ln \langle T(x)\rangle$ as a function of $X=2D^2\ln (x/\tilde x)$ for 
different values of $D$, and compare with the theoretical scaling function, given by equation (\ref{eq: 5.7}).

\section{Conclusions}
\label{sec: 6}

In his analysis of equation (\ref{eq: 1.1}), Zwanzig considered the mean-first-passsage time
$T(x)$ to reach displacement $x$. Computing the expectation value $\langle T(x)\rangle$ 
over different realisations of the random potential, he showed \cite{Zwanzig1988} that $\langle T(x)\rangle\sim x^2$, 
which is consistent with a diffusive dispersion, with an effective diffusion coefficient $D_{\rm eff}$. 
The effective diffusion coefficient vanishes in a highly singular manner as $D\to 0$, and numerical 
studies have suggested that equation (\ref{eq: 1.1}) exhibits anomalous diffusion \cite{Khoury2011,Simon2013,Goychuk2017}.
It seems evident that the discrepancy between these two pictures of the dynamics results 
from the expectation value $\langle T(x)\rangle$ being dominated by rare events, where an unusually large 
fluctuation of the potential $V(x)$  acts as a barrier to dispersion. The central limit theorem 
is applicable to this problem, and at sufficiently large values of $x$ the ratio $T(x)/\langle T(x)\rangle$ is 
expected to approach unity, for almost all realisations of $V(x)$. However, at values of $x$ which are 
of practical relevance, most realisations $T(x)$ will be much smaller than $\langle T(x)\rangle$.

In order to give a description of the dynamics of (\ref{eq: 1.1}) which is both empirically useable and 
analytically tractable, we considered the median (with respect to different realisations of the potential) 
of the mean-first-passage time. In the limit where the diffusion coefficient $D$ is small, the integrals 
which appear in the expression for the first passage time, equation (\ref{eq: 2.1}), are dominated 
by maxima and minima of the potential, described by equations (\ref{eq: 2.2.2}) and (\ref{eq: 2.2.4}).
This observation led us to consider the statistics of sums of exponentials of random variables, 
equations (\ref{eq: 1.11}) and (\ref{eq: 3.1}). We gave a quite precise estimate, equation (\ref{eq: 3.1.18}), for the 
median of (\ref{eq: 1.11}) and also derived simple relations describing the asymptotic behaviour of these 
sums, equations (\ref{eq: 3.2.6}) and (\ref{eq: 3.2.8}). 

It is these expressions which enable us to formulate a 
concise asymptotic description of the dynamics of (\ref{eq: 1.1}) in the limit as $D\to 0$, in terms of the 
large deviation rate function of the potential, $J(V)$.
We argued that at very long length scales $\bar T(x)$ approaches the expectation value $\langle T(x)\rangle$, 
and that the dispersion is diffusive, in accord with the theory of Zwanzig \cite{Zwanzig1988}. On shorter 
timescales $\bar T(x)$ is determined by a \lq flooding' model, according to which the probability density for locating the 
particle occupies a region which is constrained by a potential barrier which can trap a particle for time $\bar T$. 
As $\bar T$ increases, higher and higher barriers are required. For a Gaussian distribution of barrier heights, 
equation (\ref{eq: 4.5}) implies that the dispersion is described as sub-diffusive, of the form
\begin{equation}
\label{eq: 6.1}
x\sim \tilde x\exp\left(\frac{D^2(\ln \bar T)^2}{8C(0)}\right)
\end{equation}
which is distinctively different from the power-law anomalous diffusion 
which has been reported by some authors \cite{Khoury2011,Simon2013,Goychuk2017}.
Our numerical investigations of the dynamics of equation (\ref{eq: 1.1}) for different 
values of $D$, illustrated in figure \ref{fig: 4}, show a data collapse which is in excellent 
agreement with equation (\ref{eq: 5.7}), verifying (\ref{eq: 6.1}).

{\bf Acknowledgments}. We thank Baruch Meerson for bringing \cite{Zwanzig1988} to our notice, 
and for interesting discussion about the statistics of barrier heights.  
MW thanks the Chan-Zuckerberg Biohub for their hospitality.

\end{document}